# A Survey and Taxonomy of Distributed Data Mining Research Studies: A Systematic Literature Review

Fauzi Adi Rafrastara, *Graduate Student Member, IEEE*, Qi Deyu

*Abstract— Context:* Data Mining (DM) method has been evolving year by year and as of today there is also the enhancement of DM technique that can be run several times faster than the traditional one, called Distributed Data Mining (DDM). It is not a new field in data processing actually, but in the recent years many researchers have been paying more attention on this area. *Problems:* The number of publication regarding DDM in high reputation journals and conferences has increased significantly. It makes difficult for researchers to gain a comprehensive view of DDM that require further research. *Solution:* We conducted a systematic literature review to map the previous research in DDM field. Our objective is to provide the motivation for new research by identifying the gap in DDM field as well as the hot area itself. *Result:* Our analysis came up with some conclusions by answering 7 research questions proposed in this literature review. In addition, the taxonomy of DDM research area is presented in this paper. Finally, this systematic literature review provides the statistic of development of DDM since 2000 to 2015, in which this will help the future researchers to have a comprehensive overview of current situation of DDM.

## I. INTRODUCTION

Recently the centralized data mining techniques are commonly used to analyze the large either corporate or scientific data which stored in database [1]. The main challenge in data mining is to find the relationship among data quickly and correctly [2]. The emerging of large and big data yields the heavy process of the single computer to complete the calculation task. However such significant growth of the data volume day by day forces the researchers to provide more advanced method or strategy to solve this problem.

Over the last few years, parallel and distributed computing became more famous mainly on data processing and information extraction. The birth of distributed computing over several years ago could deal with this current problem in which the mined data currently is not only in range of Megabytes to Gigabytes, but even more than Terabytes and Petabytes. Social media and web service produce a fantastic amount of data which touching the scale of Petabytes daily. The existence of large dataset and the needs to process that information quickly makes the use of distributed or parallel computing is really important today [3].

The commodity hardware currently can be connected to the clusters easily for running the complex task in distributed environment. The combination of data mining and distributed computing can improve the mining performance of data mining algorithm especially in large and distributed dataset. Recently the emerging of DDM becomes extremely important. It focuses on the data analysis in distributed environment while paying attention on several issues related to the computation problem, storage, data communication and human-computer interaction as well [4].

This paper will discuss the current research of distributed data mining (DDM). We downloaded and reviewed 486 high quality research studies to provide the statistics, mind map, and taxonomy regarding the situation of DDM research nowadays. This paper consists of 4 chapters. First chapter is introduction. The second chapter discusses the methodology that we used. The statistical result will be shown and explained in chapter 3. Last chapter will provide the conclusion of this research.

## II. RESEARCH METHOD

For reviewing the literature on the DDM field, a systematic methodology was applied in this work. Systematic Literature Review (SLR) initially was a well-known systematic review approach in software engineering area [5][6][7][8][9] and currently becoming more popular in other computer science fields as well, such as cloud computing [10], distributed computing [11], and internet technology [12].

The main objective of SLR is to present the correct assessment, identification and interpretation of all available research evidence regarding the research topic being studied, using the reliable, rigorous and auditable methodology. Finally, SLR can answer the specific research questions based on the collected data after completing the review process [5][11][9].

The review method which applied in this work was following the guidelines proposed by Kitchenham and Charters [11], and also inspired by some other researchers

Fauzi Adi Rafrastara is with the School of Computer Science and Engineering, South China University of Technology, Guangzhou, China, 510006, on leave from Dian Nuswantoro University, Semarang, Indonesia. E-mail: fauzi_adi@yahoo.co.id.
Qi Deyu is a Professor in School of Computer Science and Engineering, South China University of Technology, Guangzhou, China, 510006. E-mail: qideyu@scut.edu.cn.

[9][5][8][12][10].

*A. Review Method*

In this work, SLR is divided into 3 stages, namely: planning, conducting and reporting the literature review. On the first stage, there are 3 steps involved. Firstly we identify the requirements for a systematic review. Step 2 is performed to develop the review protocol which is used as a foundation to obtain the sharp result and to reduce the possibility of researcher bias. In this step, research questions are constructed along with defining the search strategy, inclusion and exclusion criteria, quality assessment, and finally data extraction and synthesis process. Those all parts of review protocol are discussed in Section 2.1, 2.2, 2.3, 2.4, 2.5, 2.6. In step 3, we evaluate the developed review protocol. This evaluation is done in planning stage and improved iteratively during the conducting and reporting stage.

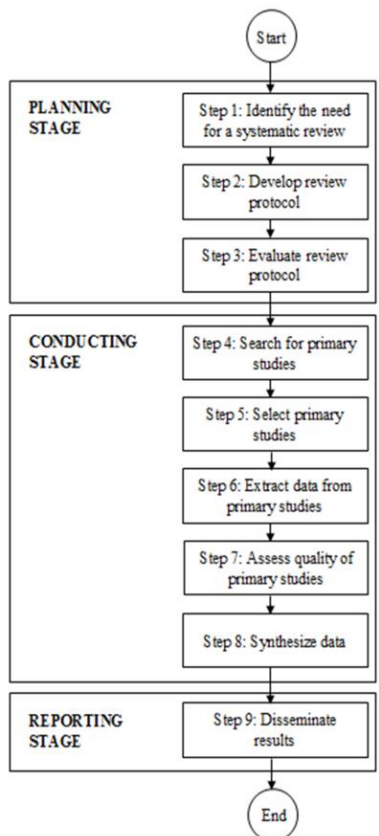

Fig. 1. The steps of SLR Process.

*B. Research Question*

Research Question (RQ) is formulated mainly to make the review stay focused. According to the PICOC criteria introduced by Kitchenham and Charters [11], we present the summary of Population, Intervention, Comparison, Outcomes, and Context of our research in Table 1.

Based on the PICOC table in Table 1, we develop the research questions and motivation in this literature review as shown in Table 2.

RQ1 to RQ3 are constructed to help researchers to evaluate the context of the primary studies. They provide the summary and synopsis of some particular publications, authors and research areas in DDM field. On the other hand, RQ4 to RQ7 are the main research questions on this literature review. They talk about the datasets, popular methods and new proposed methods in this field.

TABLE 1
SUMMARY OF PICOC

| Population | Distributed system, parallel system |
|---|---|
| Intervention | Data mining, methods, algorithms, techniques, datasets |
| Comparison | - |
| Outcomes | Successful DDM methods |
| Context | Studies in industry and academia |

TABLE 2
RESEARCH QUESTIONS AND MOTIVATIONS OF LITERATURE REVIEW

| ID | Research Question | Motivation |
|---|---|---|
| RQ1 | Which journal is the most significant in DDM journal? | Identifying the most significant journal in DDM |
| RQ2 | Who are the most active and influential researchers in the DDM field? | Identifying the most active and influential researchers who contributed so much in DDM field |
| RQ3 | What kind of research topics are selected by researchers in the DDM field? | Identifying the research topics and trends in DDM |
| RQ4 | What kinds of datasets that commonly used for DDM? | Identifying the datasets that commonly used in DDM |
| RQ5 | What kinds of methods are used for DDM? | Identifying opportunities and trends for DDM's method |
| RQ6 | What kinds of methods are the most used for DDM? | Identifying the most used methods in DDM |
| RQ7 | What kinds of method improvements are proposed for DDM? | Identifying the proposed method improvements for DDM |

To give the simpler illustration regarding our research questions, the basic mind map of this systematic literature review is provided as shown in Figure 2.

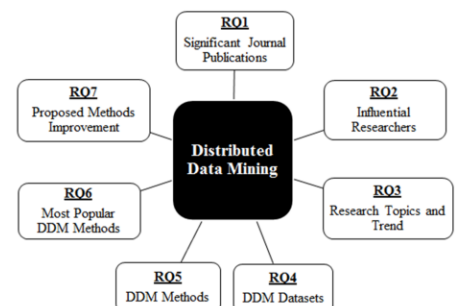

Fig. 2. Mind Map of Research Questions.

*C. Search Process*

On the conducting stage, the step is started with searching for primary studies. It consists of several activities, such as selecting digital libraries, defining the search string and retrieving the high quality papers that related to the research topic which being discussed. To find the high quality papers, some digital libraries must be specified first. Three well known literature databases in the field of computer science are selected and listed as follows:

*1) ACM Digital Library (dl.acm.org)*
*2) IEEE eXplore (ieeexplore.ieee.org)*
*3) ScienceDirect (sciencedirect.com)*

Specific search string is used to collect the articles from those three digital libraries. The search string is developed based on the following steps:

*1) Defining the search term based on PICOC criteria, especially from Population and Intervention.*
*2) Defining the search term from research questions.*
*3) Defining the search term in relevant titles, abstracts and keywords.*
*4) Defining the synonyms, antonyms and alternative spellings of search terms.*
*5) Implementing the advanced search string using identified search terms, Boolean AND and OR.*

The resulted search string is as follows:

*(Distributed OR parallel) AND (("data mining") AND (method\* OR algorithm OR datasets))*

The search string is adjusted depend on the specific requirements of each digital library. However, the original search will be kept to avoid the significant increment of irrelevant studies. During the search process, search string is implemented based on the title, abstract and keyword of the documents. The studies selected in this literature review are the high reputation journals that obtained from 3 popular online digital libraries: ScienceDirect, IEEExplore, and ACM. The search process is conducted in the end of Mei 2015, covering the papers published since January 2000 to Mei 2015

TABLE 3
INCLUSION AND EXCLUSION CRITERIA

| | |
|---|---|
| **Inclusion Criteria** | Studies that discussing data mining technique in distributed environment. |
| | Studies that discussing the improvement or implementation of DDM and conducting the experiment using at least 1 DDM algorithm. |
| | Studies published in journal, transaction or high quality conference. |
| | For the same studies that have duplicate publication, only the most complete and newest one will be included. |
| | Studies published within January 2000 to Mei 2015. |
| **Exclusion Criteria** | Studies that discussing Data Mining but not using Distributed/Parallel System. |
| | Studies that discussing Distributed/Parallel System but not using Data Mining techniques. |
| | Studies without experimental process and result using at least 1 DDM Algorithm. Demonstration product or software will not be considered as an experimental process. |
| | The data is not a text or number. Graph and Image will be excluded. |
| | Studies that not written in English |

TABLE 4
STUDIES SELECTION

| No. | Publisher | Stages | | | | |
|---|---|---|---|---|---|---|
| | | 1 | 2 | 3 | 4 | 5 |
| 1. | SD | 336 | 328 | 82 | 68 | 68 |
| 2. | ACM | 35 | 20 | 9 | 6 | 6 |
| 3. | IEEE | 97 | 33 | 14 | 11 | 11 |
| **Total** | | 468 | 381 | 105 | 85 | 85 |

*D. Paper Selection*

Primary study selections are conducted by using the inclusion and exclusion criteria of the searched articles. Table 3 is showing the accepted and unaccepted criteria of the documents being reviewed.

The search result is stored and managed using a software package, called Mendeley (http://mendeley.com). There are 5 stages in paper selection which described as follows:

*1) Applying the search query to all digital libraries.*
*2) Excluding the invalid and duplicate documents.*
*3) Applying inclusion and exclusion criteria to the papers title, abstract and keywords.*
*4) Applying inclusion and exclusion criteria to the introduction and conclusion part of the papers.*
*5) Reviewing the selected documents and applying the inclusion and exclusion criteria to the text or content.*

Table 4 is showing the stages along with digital libraries and numbers of study that has been identified. All of the involved studies are listed in Table A-4 in Appendix.

III. RESEARCH RESULT

*A. Significant Journal Publications*

Among 85 final studies which downloaded from 3 digital libraries, there are 34 journal names and 4 publishers that successfully identified. According to the Scimago Journal Ranking (SJR) (http://scimagojr.com) those papers vary in SJR's indicator and quartile category. 2 journals published more than 10 articles discussing about DDM, such as: Future Generation Computer Systems and Journal of Parallel and Distributed Computing, in which both are published by Elsevier. 17 Journals only published each 1 paper related to this topic. The detail SJR statistic of selected studies is shown in Appendix, Table A-5.

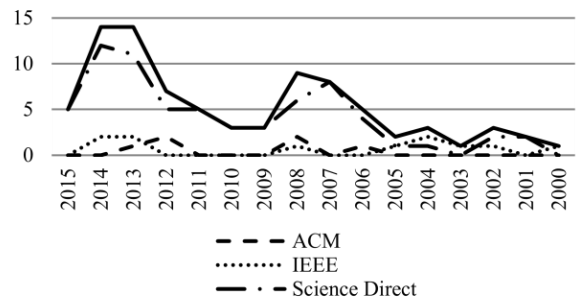

Fig. 3. The growth of studies about DDM year by year

*B. Most Active and Influential Researchers*

According to the data successfully collected, 251 researchers are involved in publishing 85 high quality papers regarding DDM. However, there are only 17 researchers which published 2 or more papers in ACM, IEEE or ScienceDirect. Fig. 4 shows the name of the researchers who published more than 1 paper in DDM area. Unfortunately, only 4 researchers which put their name as a first author and only one of them that constantly published up to 3 papers. Jaideep Vaidya and Domenico Talia can be noted that they are the most active researchers in this area with 3 publications, and surprisingly Jaideep Vaidya has successfully published his 3 papers as a first author.

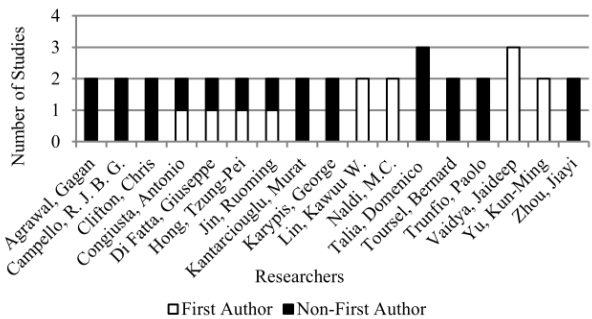

Fig. 4. Most active and influential researchers in DDM field.

### C. Research Topics in DDM

In this section, we categorize the paper contents into 3 categories, those are: Improvement, Implementation and Parallelism. Improvement means the researchers proposed the novelty or improvement in DDM area. Mostly it relates to the improvement of the existing algorithm and improvement in data security.

Implementation means the researchers attempted to apply or implement the current DDM technology to satisfy their needs. It can be the implementation of existing DDM algorithm to different area, such as computer science, medical, or even transportation.

Whereas parallelism is an effort of the researchers to convert the conventional data mining into distributed one. They modified and improved the DM algorithm to be DDM algorithm, and then compared each other to show and prove that DDM algorithm is better than conventional one.

Regarding the improvement part, we conclude that there are three areas targeted by the researchers, called: efficiency, effectiveness and security. Most of researchers are focusing on improvement of efficiency in DDM, followed by improvement on security and effectiveness.

Talking about improvement of efficiency, it contains speedup, resource & cost, and scalability. Speedup focuses on the improvement of the speed during processing data and collecting result. Resource & cost means invention of reducing the involved resource during the data processing as well as minimizing the cost for the project. Improvement of scalability is more about ideas to create a technology that can be scaled up easily according to the needs of fast data processing with much bigger data later on. As a result, 25 papers are discussing the improvement in efficiency, and 17 of them are focusing on speedup enhancement. Statistic of papers distribution regarding improvement of efficiency is illustrated in Fig. 6 and Fig. 7.

On the other hand, improvement of effectiveness in DDM research field mainly focuses on the level of accuracy during the mining process and result gathering. Information produced by DDM technology should be much better in term of accuracy, day by day, so that it can help to make a better decision as well. Not many researchers pay more attention on this field. It was only 4 studies that proposing the enhancement in term of accuracy.

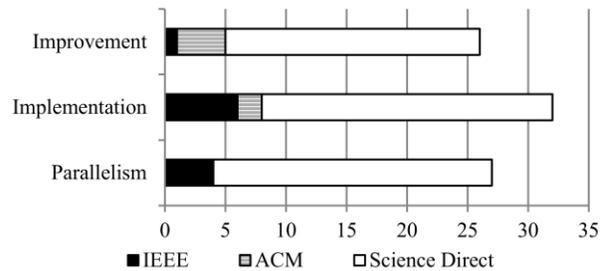

Fig. 5. DDM Research paper categories

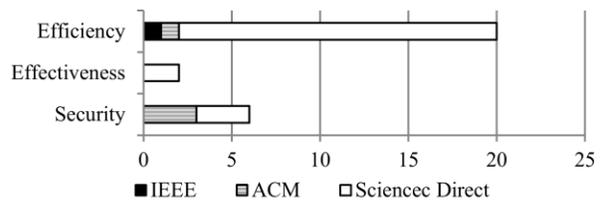

Fig. 6. Statistic of studies of improvement.

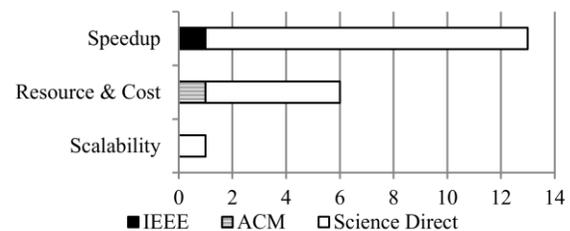

Fig. 7. Statistic of studies of improvement in efficiency.

Regarding the enhancement of security part in DDM, 9 studies are identified proposing the new level of security. All of them are discussing about privacy preserving.

In 2015, Loh & Yu [13] introduced CudaSCAN, the improvement of performance of DBSCAN algorithm by adding the power of GPU accelerator. They were not only simply adding GPU technology inside, but they enhanced the use of GPU so that it could perform better in term of efficiency (speed up) compared to CUDA-DClast, the existing GPU-based DBSCAN [14]. Cuzzocrea et al. also discussed the hottest topic here (speedup), by proposing the so called Tree-based Distributed Uncertain Frequent Itemset Mining [15]. This algorithm mainly is used to mine the constrained frequent itemsets from distributed uncertain data.

Enhancement in accuracy has been done by Di Fatta et. al. [16] in which they improved the k-means algorithm to be the so called Epidemic K-Means algorithm. It is a fully distributed K-Means method that does not require global communication and it is intrinsically fault tolerant. The authors of [17] also proposed a novel idea in security area, by using harmony search and pruning ensemble for malware detection. According to their experiment, this algorithm outperforms the existing ensemble algorithm in term of detection accuracy.

In the security field, the algorithm improvement has been done by adding the privacy preserving feature into several

DDM algorithms, such as Naive Bayes [18], ID3 [19], Random Forest [20], Apriori [21], Back Propagation [22], etc.

Since the term of data mining was firstly introduced in the computer science field in 1989 [23], so it can be normal if DDM is much more famous in computer science rather than other fields of science. From the Fig. 8, we can see that the difference between numbers of studies discussing DDM in Computer Science (CS) and other disciplines is too wide. There are 82 studies discussed purely about CS and only 2 studies about collaboration between CS and medical. There is only 1 paper involving transportation field that has been recorded in this survey.

Zheng & Wang in 2014 attempted to parallelize the Pruning Eclat Algorithm by using MapReduce to study the method of road transport management information [23]. By the parallelism, this algorithm achieved a better performance by reducing a time waste more than 40% compared to the conventional one. Parallel implementation is also used beyond the computer science field, such as Medical, wherein Genetic Algorithm was parallelized to analyze the large datasets as published by Rausch et al. in 2008. They proposed such technique to discover patterns in genetic markers that indicate the tendency to multifactorial disease [24]. Olejnik et al. also discussed the cross-field research when they implemented the parallel algorithm in medical science field. The paper involved Clustering Distributed Progressive algorithm using DiabCare Medical Database [25].

Especially in computer science area, since the number of identified studies is too big, then we break it down into 8 categories, those are: Network & Internet, Software Engineering, Security & Privacy, Hardware Acceleration, Bioinformatics, E-Government, and General Data Mining Area. Topic distribution of the DDM studies can be seen in Figure 9. Network & Internet, Security & Privacy, and Hardware Acceleration can be considered as hot topic beside General DDM field. 35 studies have been identified that their focuses are on the general DDM area. This topic covers the discussion about DDM method improvement, implementation, or parallelism without concerning to any other computer science fields.

*D. Datasets Used in DDM*

This section will discuss deeper about datasets used by researchers in DDM. According to the access of the dataset, we classify them into two categories, namely: public and private dataset. Those classifications are derived from the dataset used and mentioned by the researchers in their studies. 42% of researchers use private databases, whereas 38% use the public one. Surprisingly, 16% of studies involve both public and private datasets. The rest 4% are considered using unknown dataset, since they did not mention what dataset that they use. The detail statistic regarding dataset and research paper can be seen in Fig. 10 (*Left*).

Most of the public dataset recorded in this survey are derived from UCI machine learning (*http://archive.ics.uci.edu/ml/datasets.html*).

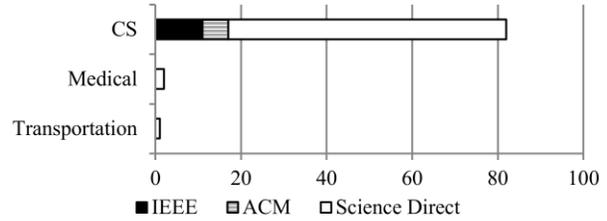

Fig. 8. Statistic of DDM studies in CS and non CS.

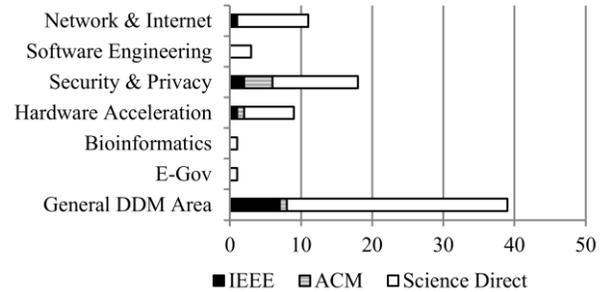

Fig. 9. Statistic of DDM studies in CS field.

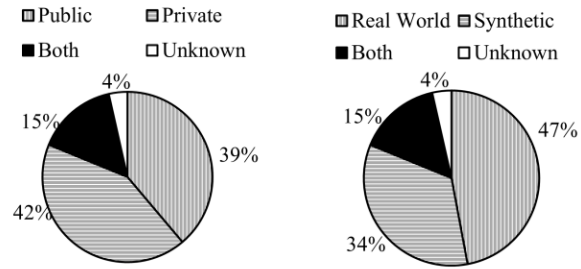

Fig. 10. Statistic dataset based on access (*Left*) and origin (*Right*)

Regarding the origin of dataset, we divide them into two groups, called: real world and synthetic dataset. Real world dataset means the researcher captured the real data directly from the nature, whereas synthetic dataset means data is obtained from the simulation or created by themself. 47% of studies use the real world datasets, whereas 34% of them use the synthetic one. What interesting is 16% of studies use both real world and synthetic data for their experiment.

Most of the public dataset that used by DDM researchers collected from the UCI Machine Learning Repository. There are 37 studies using this repository and 120 different UCI datasets have been downloaded for their experiments. Iris and Mushroom become the most popular UCI dataset since they are used in 9 different DDM papers. In addition, 9 papers with private dataset collected the data by utilizing IBM Generator Tool.

*E. Method Used in DDM*

According to Luo et al., algorithm library layer of data mining is composed by three main components, such as: association, classification, and clustering [1]. This paper is following their idea and breaking down the DDM algorithm into the same 3 main problems.

Based on this review, 85 DDM studies come up with 192 numbers of algorithms in total, in which some of them use the same popular algorithm, such as apriori, k-means, knn, and decision tree. Surprisingly, we have found 137 different kinds of methods that used in DDM research, including association, classification and clustering algorithm.

Detail statistic of methods involved is listed in Table A-2 (Appendix). The Fig. 12 illustrates the distribution of methods according to data mining categories.

We have found the fact that there are 53 different clustering algorithms, 42 classification algorithms and 42 association algorithms have been discussed in 85 reviewed research papers. It means that there are some studies involving more than one DDM problems in a single research paper. In 2013, Villar et al. did an experiment by involving genetic algorithm (classification) and hill climbing algorithm (clustering) to get the optimal internal configuration of all the switches in the network of large supercomputers that running parallel applications [26]. In another research article, Ericson and Pallickara studied 4 clustering algorithms and 2 classification algorithms by doing the benchmark test using Mahout back-end code [27].

As mentioned in the beginning of this section, 192 numbers of algorithms are involved in our 85 main research papers. Most of them are clustering algorithms, with 74 algorithms, followed by classification and association with 63 and 55 algorithms respectively. The detail statistic regarding DDM methods reviewed in this paper is listed in Appendix, Table A-2.

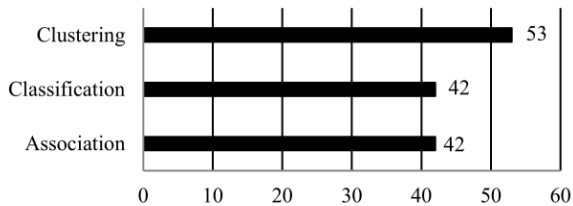

Fig. 11. Statistic of methods involved

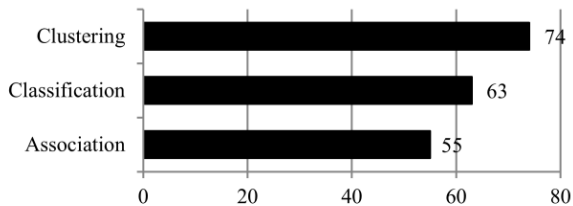

Fig. 12. Statistic of frequent methods used

*F. Most Used Method in DDM*

By using the statistic from the previous section, now we can come up with the statistic of the most used method in DDM research (*see Table 4*). In Association category, 5 algorithms are used more than once. The most popular one is Apriori algorithm with 9 papers involved.

TABLE 4
MOST USED DDM METHODS

| Category | | Algo Name | Studies | Totals |
|---|---|---|---|---|
| Association | 1. | Apriori | [14][15][29][30][31][32][33][20][34] | 9 |
| | 2. | Eclat | [28][35] | 2 |
| | 3. | FP-Tree | [36][37] | 2 |
| | 4. | TPFP | [36][38] | 2 |
| | 5. | BTPFP | [36][38] | 2 |
| Classification | 1. | ID3 | [18][39][40][41] | 4 |
| | 2. | C4.5 | [33][41][42][43][44][45][46][47][48] | 9 |
| | 3. | Neural Network | [49][50][51] | 3 |
| | 4. | Naive Bayes | [17][26][52] | 3 |
| | 5. | K-Nearest Neighbor | [37][53][54] | 3 |
| Clustering | 1. | K-Means | [37][42][26][53][54][55][56][57][58][15][59][60] | 12 |
| | 2. | P2P K-Means | [56][61][15] | 3 |
| | 3. | Expectation Maximization (EM) | [34][53][62][63][64] | 5 |
| | 4. | DBSCAN | [33] [65] | 2 |
| | 5. | SOM | [66][67] | 2 |
| | 6. | AutoClass | [42][68] | 2 |
| | 7. | DPC | [55][24] | 2 |

On the other hand, Classification category has 42 different algorithms whereby 5 of them can be regarded as the most popular one since their algorithms are involved in more than 2 research studies, i.e. ID3, C4.5, Neural Network, Naive Bayes, and K-Nearest Neighbor.

Finally in clustering area, 7 out of 53 algorithms can be noted as the most popular one. K-Means algorithm is leading conveniently with 12 studies involved, whereas Expectation Maximization (EM) and P2P K-Means has only 5 and 3 research papers respectively.

*G. Proposed Method Improvement for DDM*

Around 30% of the studies that have been reviewed are talking about DDM improvement. Most of the improvement parts are done in Association area with 15 papers involved. The distribution of improvement papers either in classification and clustering are the same, both of them have 8 studies only.

The list of new algorithms can be seen in Table A-3 (Appendix). In Association category, 14 algorithms belong to Frequent Pattern Mining, whereas Sequence Pattern Mining only has 1 proposed algorithm, called Prioritized Sensitive Patterns with Dynamic Blocking [28]. In 2010, Yu & Zhou [29] attempted to improve the capability of Parallel FP-Tree, by proposing a novel approach called Tidset-based Parallel FP-Tree (TPFP) and Balanced Tidset-based Parallel FP-Tree (BTPFP). In 2013, their proposed algorithm both are implemented on the research of Lin & Lo to construct 4 novel algorithms, they are: Equal Working Set (EWS), Request on Demand (ROD), Small Size Working Set (SSWS), and Progressive Size Working Set (PSWS) algorithm. Their improvement can provide a fast and scalable mining service for frequent pattern mining in many-task computing environment.

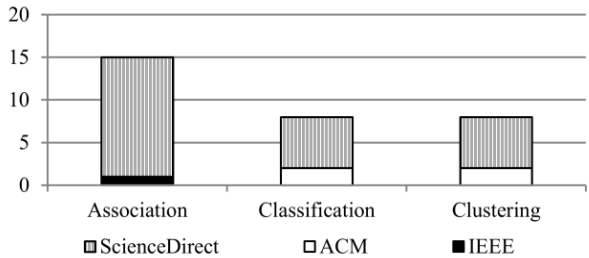

Fig. 13. Numbers of paper discussing improvement of DDM algorithm

In Classification field, there is an interesting fact that 5 out of 8 proposed methods are focused on security part. The goal of such methods is to avoid disclosing data beyond its source. The researchers proposed the hybridization of algorithm by combining the DDM classification method with privacy preserving technique. Vaidya et. al. improved the security level of ID3 Decision Tree and Naive Bayes by adding the capability of privacy preserving to those both algorithms [19][18]. In [18] they proposed that secure classification algorithms for vertically and horizontally partitioned data, whereas in [19] only for vertically partitioned data. In 2013, Sheen et al. [17] proposed a music inspired algorithm, called Harmony Search Ensemble (HS_ENSEM). This method is utilized for malware detection. An Ensemble is constructed by using multiple heterogeneous classifiers in parallel fashion. To get the pruned set, the harmony search is utilized to choose the best set of classifiers which obtained from the ensemble.

By the 8 novel clustering algorithms, K-Means becomes popular algorithm to improve as well as Expectation Maximization (EM). The original K-Means was improved twice to be Sequential Sampling Spectral K-Means [30] and Epidemic K-Means [16]. In 2012 Di Fatta et al [16] proposed the fully distributed K-Means Algorithm (Epidemic K-Means) wherein they claimed that global communication is not required anymore by using this method and it is intrinsically fault tolerant. As mentioned above, EM also becomes famous algorithm here as it is used and improved twice by the different researchers [31][32]. In 2005 the basic EM was combined with Privacy Preserving (PP) technique by Merugu and Ghosh [31] to enhance the security level of EM method. At the end, the researchers claimed that based on their experiment, PPEM can achieve the high quality global cluster with little loss of privacy. This technique actually is based on building probabilistic models of the data at each local site, in which the parameters are transmitted to a central location afterwards. Another story happened in 2015 where Loh & Yu [13] proposed an improvement in clustering algorithm. They successfully improved the efficiency of DBSCAN algorithm, by proposing a novel technique called CudaSCAN. The researchers explained that there are three phases in CudaSCAN: (1) Partitioning the entire dataset; (2) local clustering within sub-regions in parallel; (3) merging the local clustering results. By their experiments, they claimed that CudaSCAN outperformed CUDA-DClust (a previous DBSCAN extention), by up to 163.6 times.

*H. DDM Taxonomy*

By considering the review outcome from 86 high reputation studies, DDM taxonomy is proposed to map the current situation of DDM research. This taxonomy is aimed to help the future researcher to get the simple way to understand the general DDM area, including the hot area inside and the gap between topics under DDM field. This taxonomy is developed from the data collected during the review process. The references supporting this taxonomy are provided in Table A-4 (Appendix).

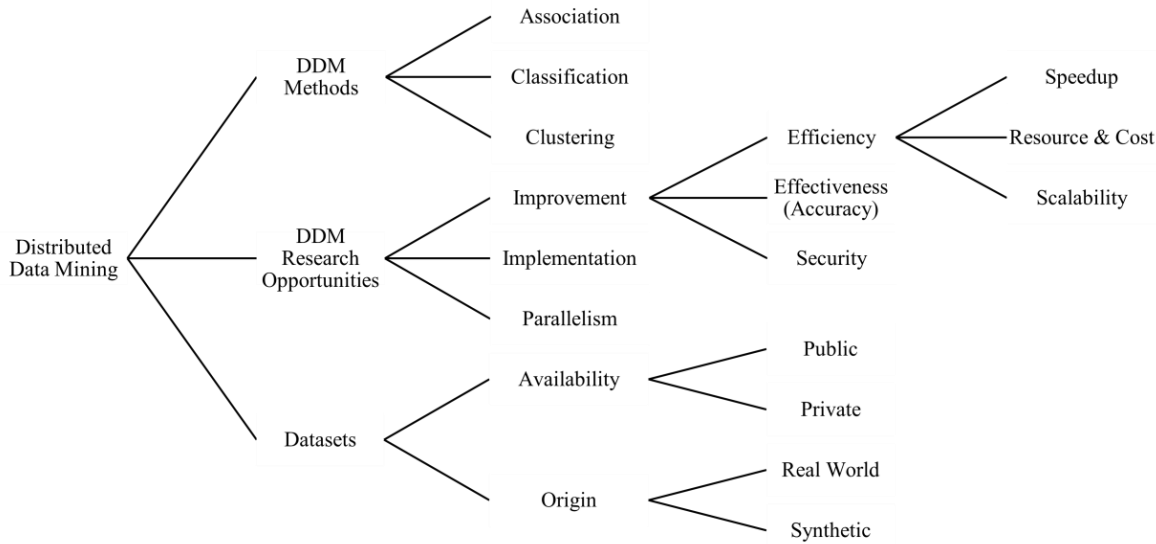

Fig 14. DDM Taxonomy

Firstly DDM is divided into three categories, those are: DDM Problem, DDM Research, and Dataset. In DDM Problem, we classify the reviewed papers into 3 sub categories, namely: Association, Classification, and Clustering. The DDM research study can be one of those items or combination of several categories that mentioned above.

DDM Research has also 3 sub categories, called: Improvement, Implementation, and Parallelism. Only Improvement that has sub categories, those are: Efficiency, Effectiveness (accuracy), and Security. Efficiency itself can be explored deeper by classifying it into 3 groups: Speed up, Resource & Cost, and Scalability. Effectiveness is emphasizing on the accuracy of the algorithm, whereas in the security part, the previous researchers just focused on the implementation or improvement of the privacy preserving in DDM field.

The Dataset can be assessed from their availability and their origin. Availability means weather the dataset is publicly available or not. Origin means weather the researcher captures the dataset from the real data or generates it synthetically. The complete DDM Taxonomy is illustrated in Fig. 14.

## IV. CONCLUSION

This survey is conducted using systematic literature review methodology. The aim of SLR is to find the answers of some specific research questions by analyzing, assessing, identifying and interpreting all available research evidences related to the research topic being studied, using the reliable, rigorous and auditable methodology [5][9][11].

The contribution of SLR in this study is mainly to identify and analyze the trends, datasets and methods used in DDM research between 2000 and 2015. By following the SLR methodology, we collected more than 400 high quality journal articles from three major digital libraries (such as: IEEE, ScienceDirect, and ACM), and finally 85 papers that have direct discussion about DDM were selected. Seven research questions has been constructed, explored, and answered in this study as well.

Based on the analysis of the selected studies, those papers can be categorized into 3 major focuses: DDM Research Opportunities, DDM Datasets, and DDM Methods. Regarding DDM Research Opportunities, actually DDM area is still widely opened for all researchers around the world, since the research improvement that has been done is not so high, it is only 26 out of 85 DDM journal papers, or about 30.6 % from the total selected studies. In addition, the gap between improvement in efficiency and effectiveness is quite big. There are 20 papers proposing a new method for efficiency, but only 2 papers emphasizing on the enhancement of effectiveness. By this fact, the effectiveness of DDM technique, especially about level of accuracy, can be an interesting topic for the next DDM research, so that the quality improvement of a new method is not merely about speed or efficiency, but the accuracy as well. The improvement of security is also cannot be underestimated since the privacy preserving issue will always be the hot issue to be discussed and improved.

The majority implementations of DDM are in computer science field. It is noted that there are only 2 papers discussing DDM in the area beyond computer science according this survey, those are 2 papers in Medical field and 1 paper in Transportation field. It means that the area of DDM research actually is wide and open. A lot of research field that have not been influenced by DDM technology yet. Or if there are some other fields that already used DDM technology, but their research and innovation have not been published yet. It can be a good opportunity for the researcher to do such kind of research and publish it.

On the other hand, we have found a fact that majority papers in DDM field are using the private dataset. Actually the proportion of private and public dataset according to our survey is almost the same. However the numbers of private dataset that have been used by DDM researchers is slightly higher than the public one. There are 42% of DDM papers are using the private dataset and 39% of them are using the public data. And surprisingly there are 15% of the studies that use both private and public dataset. However, the fact about the dominance of the use of the private dataset is very lamentable. It can be a critical problem for the continuity of DDM research later on, since a proposed method cannot be compared with the existing method if the dataset is private. It is impossible to make sure whether the result of the proposed method is surely better than the existing one or not.

Regarding the method used in DDM field, this paper categorized it into 2 topic areas, which are: the methods mostly used in DDM research and the new methods proposed by DDM researchers. DDM method basically is coming from the Data Mining paradigm. There are three major groups on DDM method, called: Association, Classification, and Clustering. In Association area, Apriori algorithm is the most used method with 9 papers involved. C4.5 Decision Tree becomes the most famous method in Classification area with 9 papers. Whereas for the clustering algorithm, K-Means takes the place with 12 studies that conducting experiment with this algorithm.

Finally, the improvement of algorithm in DDM field since January 2000 to Mei 2015 have discussed in 26 studies (downloaded from ACM, IEEE, and ScienceDirect). Association algorithm becomes the top choice to improve in which 15 studies are discussing about improvement in Association Rule Mining. 8 improvements are happened in classification algorithm wherein security is very dominant here. 5 out of 8 studies addressed the security problem in classification algorithm. 8 novel approaches are also proposed in clustering algorithm. In this area, the enhancement of k-means and EM looks like the favorite of the researchers.

The last but not least, DDM taxonomy has been proposed with the aim to help the future researcher to have simpler and better understanding regarding DDM field. The taxonomy illustrates the map inside the DDM area so that the researcher can easily find the specific area that want to be focused to.

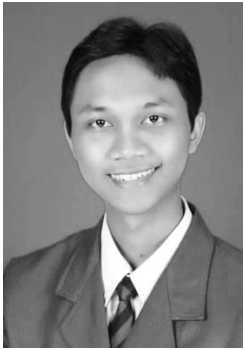
**Fauzi Adi Rafrastara** received Bachelor and Master degree in Computer Science from Dian Nuswantoro University (2009) and Technical University of Malaysia Malacca (2011), respectively. He is currently pursuing the Ph.D degree at the School of Computer Science and Engineering of South China University of Technology. He published 8 books in Indonesia and Malaysia, and several papers in International Conferences and Journals. His research interest includes data processing, multimedia, and information security. He is a member of TheIRED, the IEEE and the IEEE Computer Society.

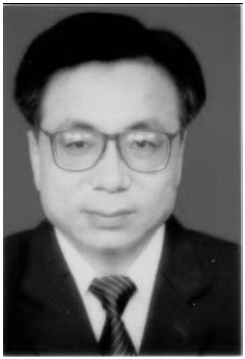
**Qi Deyu** was born in Helin county of Inner Mongolia in October, 1959. He has got bachelor degree of science, master degree of engineering, and doctor degree of engineering. Now he serves in South China University of Technology (SCUT) as the professor, tutor, leader of the academic team "advanced computing architecture" of the Xinghua engineering project, and director of the Computer Systems Research Institute of SCUT. His research area includes the computer architecture, distributed systems, computer security, etc.

**APPENDIX**

TABLE A-1

STATISTIC OF PAPERS WITH RESEARCH IMPROVEMENT

| Improvement | Focus | IEEE | ACM | SD |
|---|---|---|---|---|
| Efficiency | Speedup | [33] | | [15][34][35][36][13][20][37][38][39][29][32] |
| | Resource & Cost | | [30] | [31][40][35][41][39] |
| | Scalability | | | [40] |
| Effectiveness | Accuracy | | | [16][42][17][32] |
| Security | Privacy Preserving | | [43][19][18] | [44][28][20][31][22][21] |

TABLE A-2

THE LIST OF METHODS USED IN DDM

| Category | Methods | Studies | Methods | Studies |
|---|---|---|---|---|
| Association | Eclat | [39] [45] | Ye & Chiang Algo | [48] |
| | Apriori | [46][47][21][48][49][50][39][33][51] | EDMA | [48] |
| | Apriori + PP | [21] | P2P FISM | [40] |
| | Apriori with IDD + HD | [33] | Majority Rule | [40] |
| | U-Apriori | [15] | EWS | [36] |
| | UF-Growth | [15] | ROD | [36] |
| | FP-Growth | [39] | SSWS | [36] |
| | FP-Tree | [29] [52] | PSWS | [36][35] |
| | TPFP | [29][36] | SABMA | [55] |
| | BTPFP | [29][36] | PP-Tree | [55] |
| | TDM | [53] | LFP-Tree | [55] |
| | DPM | [53] | DFP | [55] |
| | Count Distribution | [54] | Genetic-Fuzzy | [56] |
| | PPS | [54] | PMCAR | [37] |
| | Parallel SPADE | [57] | PShrFSP-Tree | [41] |
| | PISA | [28] | ShrIO-Tree | [41] |
| | TTPC | [28] | ShaFEM | [39] |
| | STPF | [58] | FP-Array | [39] |
| | DTPF | [58] | Parallel NEclat | [23] |
| | CAR-Miner | [37] | FLR-Mining | [35] |
| | GSP | [59] | DAN-Mining | [35] |

TABLE A-2 (CONTINUED)
THE LIST OF METHODS USED IN DDM

| Category | Methods | Studies | Methods | Studies |
|---|---|---|---|---|
| Classification | Naive Bayes | [60][27][18] | Back Propagation | [22] |
| | Naive Bayes + PP | [18] | PPBP | [22] |
| | Complementary Bayes | [27] | ELM | [22] |
| | ID3 | [61][19][62][63] | PPELM | [22] |
| | ID3 + PP | [19] | PRISM | [81] |
| | C4.5 Decision Tree | [64][46][65][66][67][68][69][70][62] | MOGBAP | [82] |
| | C4.5 with Heuristic | [65] | Bagging | [17] |
| | C5.0 Decision Tree | [68] | Adaboost | [17] |
| | J4.8 Decision Tree | [71][72] | Random Subspace | [17] |
| | Decision Tree Construction | [52] | Stacking | [17] |
| | Random Decision Tree | [63] | Hill Climbing | [26] |
| | DMDT | [73] | Fuzzy Inference System | [83] |
| | Neural Network | [72][74][75] | M5P | [84] |
| | k-nearest neighbor | [76][77][52] | PRMCLP | [85] |
| | SVM | [78][79] | LVQ3 | [74] |
| | NOW G-Net | [80] | FCNN | [74] |
| | PART | [71] | DROP3 | [74] |
| | LDT | [66] | RSP3 | [74] |
| | Genetic Algorithm | [24][26] | PSO + SVM | [86] |
| | Random Forest (RF) | [20][17] | EDGAR | [38] |
| | PPRF (Random Forest) | [20] | REGAL | [38] |
| Clustering | k-means | [34][76][87][16][27][88][89][90][30][77][68][52] | Sequential Samping Spectral k-means | [30] |
| | Sequential Algorithm | [43] | CDC –sl | [97] |
| | Seq. Algorithm+PP | [43] | CDC –sl (U) | [97] |
| | Mondrian | [43] | CDC –sl SS | [98] |
| | Hilbert-curve | [43] | CDC –sl VRC | [98] |
| | HCA | [26] | CDC –al | [97] |
| | P2P k-means | [91][16][30] | CDC –al (U) | [97] |
| | Spectral k-means | [30] | CDC –al SS | [98] |
| | DBCURE-MR | [93] | CDC –al VRC | [98] |
| | Epidemic k-means | [16] | CDC –FEAC | [97] |
| | Random P2P k-means | [16] | CDC –FEAC (U) | [97] |
| | Fuzzy k-means | [27] | CDC –FEAC SS | [98] |
| | SSDSC | [30] | CDC –FEAC VRC | [98] |
| | Spectral Clustering | [30] | CDC –FEAC (10g) | [97] |
| | Intelligent Miner | [68] | CDC –FEAC (10g) (U) | [97] |
| | AutoClass | [68][92] | CDC –FEAC (10g) SS | [98] |
| | DBSCAN | [46][93] | CDC –FEAC (10g) VRC | [98] |
| | SOM | [94][42] | DF –EAC | [97] |
| | WR PSOM | [42] | DF –EAC SS | [98] |
| | EM | [31][95][76][32][49] | DF –EAC VRC | [98] |
| | KDEC-S | [44] | DF –EAC (P) SS | [98] |
| | PARC | [96] | DF –EAC VRC (P) | [98] |
| | Agglomerative Methods | [34] | DBSCAN+CudaSCAN | [13] |
| | DPC | [34][25] | DBSCAN+Cuda_DClust | [13] |
| | GMM | [32] | DBSCAN-GRID-MR | [93] |
| | Dirichlet | [27] | DBCURE-GRID-MR | [93] |
| | LDA | [27] | | |

TABLE A-3

THE NEW OR IMPROVED ALGORITHM PROPOSED IN DDM STUDIES

| Categories | New Methods | Papers | | Papers |
|---|---|---|---|---|
| Association | Apriori with IDD+HD | [27] | Privacy Preserving Distributed Association Rule Mining | [20] |
| | Equal Working Set (EWS) | [38] | ShaFEM | [28] |
| | Request On Demand (ROD) | [38] | Tidset-based Parallel FP-Tree (TPFP) | [36] |
| | Small Size Working Set (SSWS) | [38] | Balanced Tidset-based Parallel FP-Tree (BTPFP) | [36] |
| | Progressive Size Working Set (PSWS) | [38] | Tree-based DUFIM | [14] |
| | PMCAR | [71] | PShrFSP | [74] |
| | P2P-FISM | [73] | FLR-Mining | [70] |
| | Prioritized Sensitive Patterns with Dynamic Blocking | [69] | | |
| Classification | PPID3 | [18] | PPELM | [21] |
| | PPNB | [17] | PPRF | [19] |
| | EDGAR | [72] | WR PSOM | [67] |
| | PPBP | [21] | Harmony Search Ensemble (HS_ENSEM) | [16] |
| Clustering | Sequential sampling spectral k-means | [56] | Advanced EM | [64] |
| | PPSA | [75] | Epidemic K-Means | [15] |
| | KSEC-S (Security) | [76] | Distributed Progressive Clustering | [55] |
| | Advanced EM with Privacy Perserving | [62] | CudaSCAN | [13] |

TABLE A-4

THE LIST OF PRIMARY STUDIES IN THE FIELD OF DDM

| Years | Studies | Publications | | | Datasets | | | | | | Categories | | | Research Area | | | | | | | | | Algo | | |
|---|---|---|---|---|---|---|---|---|---|---|---|---|---|---|---|---|---|---|---|---|---|---|---|---|---|
| | | IEEE | ACM | SD | Pbl | Prv | N/a | Real | Synt | N/a | Ipv | Ipl | Prl | Computer Science | | | | | | | MD | TP | Asc | Clf | Clt |
| | | | | | | | | | | | | | | NI | SE | SP | HA | BI | EG | GA | | | | | |
| 2000 | [33] | ✓ | | | | ✓ | | | ✓ | | ✓ | | | | | | | | | ✓ | | | ✓ | | |
| 2001 | [64] | | | ✓ | ✓ | | | ✓ | | | | | ✓ | | | ✓ | | | | | | | | | ✓ | |
| | [57] | | | ✓ | | ✓ | | | ✓ | | | | ✓ | | | ✓ | | | | | | | | ✓ | | |
| 2002 | [80] | ✓ | | | ✓ | | | ✓ | | | | | ✓ | | | | | | | | ✓ | | | | ✓ | |
| | [46] | | | ✓ | | ✓ | | | ✓ | | | ✓ | | | | ✓ | | | | | | | | ✓ | ✓ | ✓ |
| | [94] | | | ✓ | ✓ | | | ✓ | | | | | ✓ | ✓ | | | | | | | | | | | | ✓ |
| 2003 | [92] | ✓ | | | | ✓ | | | ✓ | | | | ✓ | | | | | | | | ✓ | | | | | ✓ |
| 2004 | [68] | ✓ | | | | ✓ | | ✓ | | | | ✓ | | | | | | | | | ✓ | | | | ✓ | ✓ |
| | [58] | | | ✓ | | ✓ | | | ✓ | | | | ✓ | | | | | | | | ✓ | | | ✓ | | |
| | [51] | ✓ | | | | ✓ | | | ✓ | | | ✓ | | | | | | | | | ✓ | | | ✓ | | |
| 2005 | [31] | | | ✓ | ✓ | ✓ | | ✓ | ✓ | | ✓ | | | | | ✓ | | | | | | | | | | ✓ |
| | [52] | ✓ | | | | ✓ | | | ✓ | | | | ✓ | | | | | | | | ✓ | | | ✓ | ✓ | ✓ |
| 2006 | [91] | | | ✓ | ✓ | | | ✓ | | | | | ✓ | ✓ | | | | | | | | | | | | ✓ |
| | [44] | | | ✓ | ✓ | | | ✓ | | | ✓ | | | | | ✓ | | | | | | | | | | ✓ |
| | [53] | | | ✓ | ✓ | | | ✓ | | | | | ✓ | | | | | | | | ✓ | | | ✓ | | |
| | [77] | | ✓ | | ✓ | ✓ | | ✓ | ✓ | | | ✓ | | | | ✓ | | | | | | | | | ✓ | ✓ |
| | [54] | | | ✓ | ✓ | ✓ | | ✓ | ✓ | | | | ✓ | | | | | | | | ✓ | | | ✓ | | |
| 2007 | [61] | | | ✓ | ✓ | | | ✓ | | | | ✓ | | | | ✓ | | | | | | | | | | ✓ |
| | [96] | | | ✓ | ✓ | | | ✓ | | | | | ✓ | | | | | | | | ✓ | | | | | ✓ |
| | [34] | | | ✓ | | ✓ | | ✓ | | | ✓ | | | | | | | | | | ✓ | | | | | ✓ |
| | [65] | | | ✓ | | ✓ | | ✓ | | | | | ✓ | | | | | | | | ✓ | | | | ✓ | |
| | [47] | | | ✓ | ✓ | | | ✓ | | | | ✓ | | | | | | | | | ✓ | | | ✓ | | |
| | [71] | | | ✓ | ✓ | | | ✓ | | | | ✓ | | | ✓ | | | | | | | | | | ✓ | |
| | [21] | | | ✓ | | | ✓ | | | ✓ | ✓ | | | | | ✓ | | | | | | | | ✓ | | |
| | [66] | | | ✓ | ✓ | | | ✓ | | | | | ✓ | | | | | | | | ✓ | | | | ✓ | |

***Abbrvs:*** *Pbl=Public, Prv=Private; N/a=Not Available; Real=Real World; Synt=Synthetic; Ipv=Improvement; Ipl=Implementation; Prl=Parallelism; NI=Network & Internet; SE=Software Engineering; SP=Security & Privacy; HA=Hardware Acceleration; BI=Bio Informatics; EG=E-Government; GA=General Area of DDM; MD=Medical; TP=Transportation; Asc=Associatioin; Clf=Classification; Clt=Clustering.*

TABLE A-4 (CONTINUED)

THE LIST OF PRIMARY STUDIES IN THE FIELD OF DDM

| Years | Studies | Publications | | | Datasets | | | | | | Categories | | | Research Area | | | | | | | Algo | | | | |
|---|---|---|---|---|---|---|---|---|---|---|---|---|---|---|---|---|---|---|---|---|---|---|---|---|---|
| | | IEEE | ACM | SD | Pbl | Prv | N/a | Real | Synt | N/a | Ipv | Ipl | Prl | Computer Science | | | | | | | MD | TP | Asc | Clf | Clt |
| | | | | | | | | | | | | | | NI | SE | SP | HA | BI | EG | GA | | | | | |
| 2008 | [95] | | | ✓ | ✓ | | | ✓ | | | | ✓ | | ✓ | | | | | | | | | | | ✓ |
| | [76] | | | ✓ | | ✓ | | | ✓ | | | | ✓ | ✓ | | | | | | | | | | | ✓ |
| | [86] | | | ✓ | | ✓ | | | ✓ | | | | ✓ | | | | | | | ✓ | | | | ✓ | |
| | [79] | ✓ | | | ✓ | | | ✓ | | | | | ✓ | ✓ | | | | | | | | | | ✓ | |
| | [24] | | | ✓ | | ✓ | | ✓ | ✓ | | | | ✓ | | | | | | | | ✓ | | | ✓ | |
| | [72] | | | ✓ | ✓ | ✓ | | ✓ | | | | ✓ | | ✓ | | | | | | | | | | ✓ | |
| | [78] | | | ✓ | | ✓ | | ✓ | | | | ✓ | | | | ✓ | | | | | | | | ✓ | |
| | [19] | | ✓ | | ✓ | | | ✓ | | | ✓ | | | | | ✓ | | | | | | | | ✓ | |
| | [18] | | ✓ | | | ✓ | | | ✓ | | ✓ | | | | | ✓ | | | | | | | | ✓ | |
| 2009 | [20] | | | ✓ | | ✓ | | | ✓ | | ✓ | | | | | ✓ | | | | | | | | ✓ | |
| | [25] | | | ✓ | | ✓ | | ✓ | | | | ✓ | | | | | | | | | ✓ | | ✓ | | ✓ |
| | [73] | | | ✓ | ✓ | | | ✓ | | | | ✓ | | | | | | | | ✓ | | | | ✓ | |
| 2010 | [60] | | | ✓ | | | ✓ | | | ✓ | | ✓ | | | | ✓ | | | | | | | | ✓ | |
| | [29] | | | ✓ | | ✓ | | | ✓ | | ✓ | | | | | | | | | ✓ | | ✓ | | | |
| | [48] | | | ✓ | | ✓ | | | ✓ | | | | ✓ | | | | | | | ✓ | | ✓ | | | |
| 2011 | [67] | | | ✓ | ✓ | | | ✓ | | | | ✓ | | | | | | | | ✓ | | | | ✓ | |
| | [42] | | | ✓ | ✓ | | | ✓ | | | ✓ | | | | | | | ✓ | | | | | | | ✓ |
| | [38] | | | ✓ | ✓ | | | ✓ | | | ✓ | | | | | | | | | ✓ | | | | ✓ | |
| | [32] | | | ✓ | | ✓ | | | ✓ | | ✓ | | | | | | | | | ✓ | | | | | ✓ |
| | [87] | | | ✓ | | ✓ | | ✓ | | | | | ✓ | | | ✓ | | | | | | | | | ✓ |
| 2012 | [49] | | | ✓ | ✓ | | | ✓ | | | ✓ | | | | | | | ✓ | | | | ✓ | | | ✓ |
| | [28] | | | ✓ | ✓ | | | | ✓ | | ✓ | | | | | ✓ | | | | | | ✓ | | | |
| | [30] | | ✓ | | ✓ | | | ✓ | | | ✓ | | | | | | | | | ✓ | | | | | ✓ |
| | [22] | | | ✓ | ✓ | ✓ | | ✓ | | | ✓ | | | | | ✓ | | | | | | | | ✓ | |
| | [81] | | | ✓ | ✓ | | | ✓ | | | | ✓ | | | | | | | | ✓ | | | | ✓ | |
| | [43] | | ✓ | | ✓ | | | ✓ | | | ✓ | | | | | ✓ | | | | | | | | | ✓ |
| | [59] | | | ✓ | | ✓ | | ✓ | | | | | ✓ | | | | | | | ✓ | | ✓ | | | |

TABLE A-4 (*Continued*)

THE LIST OF PRIMARY STUDIES IN THE FIELD OF DDM

| Years | Studies | Publications | | | Datasets | | | | | | Categories | | | Research Area | | | | | | | Algo | | | | |
|---|---|---|---|---|---|---|---|---|---|---|---|---|---|---|---|---|---|---|---|---|---|---|---|---|---|
| | | | | | | | | | | | | | | Computer Science | | | | | | | | | | | |
| | | IEEE | ACM | SD | Pbl | Prv | N/a | Real | Synt | N/a | Ipv | Ipl | Prl | NI | SE | SP | HA | BI | EG | GA | MD | TP | Asc | Clf | Clt |
| 2013 | [82] | | | ✓ | ✓ | | | ✓ | | | | | ✓ | | | | ✓ | | | | | | | ✓ | |
| | [16] | | | ✓ | | ✓ | | | ✓ | | ✓ | | | ✓ | | | | | | | | | | | ✓ |
| | [27] | | | ✓ | ✓ | | | ✓ | | | | ✓ | | | | | | | | ✓ | | | | ✓ | ✓ |
| | [40] | | | ✓ | | ✓ | | | ✓ | | ✓ | | | ✓ | | | | | | | | | ✓ | | |
| | [69] | ✓ | | | ✓ | | | ✓ | | | | ✓ | | | | | | | | ✓ | | | | ✓ | |
| | [88] | | | ✓ | | ✓ | | | ✓ | | | ✓ | | | | | | | | ✓ | | | | | ✓ |
| | [89] | | | ✓ | | ✓ | | ✓ | | | | ✓ | | | | | | | | ✓ | | | | | ✓ |
| | [36] | | | ✓ | | ✓ | | | ✓ | | ✓ | | | | | | | | | ✓ | | | ✓ | | |
| | [62] | ✓ | | | ✓ | ✓ | | ✓ | ✓ | | | ✓ | | | | | ✓ | | | | | | | ✓ | |
| | [17] | | | ✓ | | ✓ | | ✓ | | | ✓ | | | | | ✓ | | | | | | | | ✓ | |
| | [55] | | | ✓ | ✓ | | | ✓ | | | | ✓ | | | | | | | | ✓ | | | ✓ | | |
| | [26] | | | ✓ | | ✓ | | | ✓ | | | ✓ | | ✓ | | | | | | | | | | ✓ | ✓ |
| | [50] | | | ✓ | | ✓ | | | ✓ | | | | ✓ | | | | | | | ✓ | | | ✓ | | |
| | [45] | | ✓ | | | ✓ | | | ✓ | | | ✓ | | | | | ✓ | | | | | | ✓ | | |
| 2014 | [15] | | | ✓ | ✓ | ✓ | | ✓ | ✓ | | ✓ | | | | | | | | | ✓ | | | ✓ | | |
| | [83] | | | ✓ | | | ✓ | | | ✓ | | ✓ | | ✓ | | | | | | | | | | ✓ | |
| | [84] | | | ✓ | ✓ | | | ✓ | | | | ✓ | | | | | ✓ | | | | | | | ✓ | |
| | [56] | | | ✓ | | ✓ | | | ✓ | | | | ✓ | | | | | | | ✓ | | | ✓ | | |
| | [93] | | | ✓ | | ✓ | | ✓ | ✓ | | | | ✓ | | | | | | | ✓ | | | | | ✓ |
| | [70] | ✓ | | | ✓ | | | ✓ | | | | ✓ | | | | ✓ | | | | | | | ✓ | | |
| | [97] | | | ✓ | ✓ | ✓ | | ✓ | ✓ | | | | ✓ | | | | | | | ✓ | | | | | ✓ |
| | [37] | | | ✓ | ✓ | ✓ | | ✓ | ✓ | | ✓ | | | | | | | | | ✓ | | | ✓ | ✓ | |
| | [85] | | | ✓ | ✓ | ✓ | | ✓ | ✓ | | | | ✓ | | | | | | | ✓ | | | ✓ | | |
| | [41] | | | ✓ | ✓ | ✓ | | ✓ | ✓ | | ✓ | | | ✓ | | | | | | | | | ✓ | | |
| | [74] | | | ✓ | ✓ | | | ✓ | | | | ✓ | | | | | | | | ✓ | | | ✓ | | |
| | [63] | ✓ | | | ✓ | | | ✓ | | | | ✓ | | | | ✓ | | | | | | | ✓ | | |
| | [39] | | | ✓ | ✓ | | | ✓ | | | ✓ | | | | | | ✓ | | | | | | ✓ | | |
| | [23] | | | ✓ | | ✓ | | ✓ | | | | | ✓ | | | | | | | | | ✓ | ✓ | | |
| 2015 | [90] | | | ✓ | ✓ | | | ✓ | | | | ✓ | | | | ✓ | | | | | | | | | ✓ |
| | [75] | | | ✓ | ✓ | | | ✓ | | | | ✓ | | | | ✓ | | | | | | | | ✓ | |
| | [35] | | | ✓ | ✓ | | | ✓ | | | ✓ | | | | | | | | ✓ | | | | ✓ | | |
| | [13] | | | ✓ | ✓ | ✓ | | ✓ | ✓ | | ✓ | | | | | ✓ | | | | | | | | | ✓ |
| | [98] | | | ✓ | ✓ | ✓ | | | ✓ | | | ✓ | | | | ✓ | | | | | | | | | ✓ |

TABLE A-5

SCIMAGO JOURNAL RANK OF SELECTED STUDIES

| No. | Journal's Name | Studies | SJR (2014) | Q Category |
|---|---|---|---|---|
| 1. | ACM Transactions on Database Systems | 1 | 1,729 | Q1 (Information Systems) |
| 2. | ACM Transactions on Knowledge Discovery from Data | 2 | 2,112 | Q1 (Computer Science (Miscellaneous)) |
| 3. | ACM Transactions on Reconfigurable Technology and Systems | 1 | 0,401 | Q2 (Computer Science (Miscellaneous)) |
| 4. | IEEE Transactions on Computers | 1 | 1,293 | Q1 (Hardware and Architecture; Software; Theoretical Computer Science) Q2 (Computational Theory and Mathematics) |
| 5. | IEEE Transactions on Cybernetics | 1 | 1,560 | Q1 (Software; Computer Science Applications; Human-Computer Interaction; Information Systems; Control and Systems Engineering; Electrical and Electronic Engineering) |
| 6. | IEEE Transactions on Dependable and Secure Computing | 2 | 1,874 | Q1 (Electrical and Electronic Engineering) |
| 7. | IEEE Transactions on Evolutionary Computation | 1 | 4,407 | Q1 (Computational Theory and Mathematics; Software) Q1 (Theoretical Computer Science) |
| 8. | IEEE Transactions on Knowledge and Data Engineering | 3 | 3,023 | Q1 (Computational Theory and Mathematics; Computer Science Applications; Information Systems) |
| 9. | IEEE Transactions on Neural Networks | 1 | | - |
| 10 | IEEE Transactions on Parallel and Distributed Systems | 1 | 2,017 | Q1 (Computational Theory and Mathematics; Hardware and Architecture; Signal Processing) |
| 11. | IEEE Transactions on Systems, Man, and Cybernetics, Part B: Cybernetics | 1 | 3,280 | Q1 (Human-Computer Interaction; Information Systems; Software; Computer Science Applications; Electrical and Electronic Engineering; Medicine (Miscellaneous); Control and Systems Engineering) |
| 12. | Ad Hoc Networks | 1 | 1,197 | Q1 (Computer Networks and Communication; Hardware and Architecture) Q2 (Software) |
| 13. | Applied Soft Computing | 2 | 2,220 | Q1 (Software) |
| 14. | Computers and Security | 1 | 1,051 | Q1 (Computer Science (Miscellaneous); Law) |
| 15. | Computers in Biology and Medicine | 1 | 0,474 | Q2 (Computer Science Applications; Health Informatics) |
| 16. | Data and Knowledge Engineering | 4 | 1,181 | Q1 (Information Systems and Management) |
| 17. | Engineering Applications of Artificial Intelligence | 2 | 1,525 | Q1 (Artificial Intelligence; Control and Systems Engineering; Electrical and Electronic Engineering) |
| 18. | Expert Systems with Applications | 7 | 1,996 | Q1 (Artificial Intelligence; Computer Science Applications; Engineering (Miscellaneous)) |
| 19. | Future Generation Computer Systems | 11 | 2,164 | Q1 (Computer Networks and Communications; Hardware and Architecture; Software) |
| 20. | Information Processing Letters | 1 | 0,904 | Q2 (Computer Science Applications; Information Systems; Signal Processing; Theoretical Computer Science) |
| 21. | Information Sciences | 4 | 3,286 | Q1 (Theoretical Computer Science; Computer Science Applications; Artificial Intelligence; Software; Information Systems and Management; Control and Systems Engineering) |
| 22. | Information Systems | 1 | 1,867 | Q1 (Hardware and Architecture; Information Systems; Software) |
| 23. | Journal of Computational Science | 1 | 0,848 | Q1 (Computer Science (Miscellaneous); Modeling and Simulation; Theoretical Computer Science) |
| 24. | Journal of Network and Computer Applications | 1 | 1,537 | Q1 (Computer Networks and Communications; Computer Science Applications; Hardware and Architecture) |
| 25. | Journal of Parallel and Distributed Computing | 10 | 1,093 | Q1 (Hardware and Architecture; Computer Networks and Communications) Q2 (Software; Artificial Intelligence; Theoretical Computer Science) |

TABLE A-5 (CONTINUED)
SCIMAGO JOURNAL RANK OF SELECTED STUDIES

| No. | Journal's Name | Studies | SJR (2014) | Q Category |
|---|---|---|---|---|
| 26. | Journal of Systems and Software | 2 | 1,381 | Q1 (Hardware and Architecture; Information Systems; Software) |
| 27. | Knowledge-Based Systems | 3 | 2,190 | Q1 (Artificial Intelligence; Information Systems and Management; Management Information Systems; Software) |
| 28. | Neurocomputing | 5 | 1,211 | Q1 (Computer Science Applications) Q2 (Artificial Intelligence; Cognitive Neuroscience) |
| 29. | Parallel Computing | 3 | 1,232 | Q1 (Software; Theoretical Computer Science; Computer Networks and Communications; Hardware and Architectures; Computer Graphic and Computer-Aided Design) Q2 (Artificial Intelligence) |
| 30. | Pattern Recognition | 1 | 2,477 | Q1 (Artificial Intelligence; Computer Vision and Pattern Recognition; Signal Processing; Software) |
| 31. | Pattern Recognition Letters | 3 | 1,294 | Q1 (Computer Vision and Pattern Recognition; Signal Processing; Software) Q2 (Artificial Intelligence) |
| 32. | Procedia - Social and Behavioral Sciences | 1 | 0,156 | - |
| 33. | Procedia Computer Science | 3 | - | - |
| 34. | VLDB Journal | 2 | 2,558 | Q1 (Hardware and Architecture; Information Systems) |
| | **Total** | **85** | | |